\documentclass[useAMS,usenatbib]{mn2e} 
\usepackage{graphicx,epsfig,color} 
\usepackage{amssymb,amsmath} 
\usepackage{pstricks} 
\newcommand{\msun}{{\rm M}_{\odot}}

\title[Triplets of SMBHs: Astrophysics, GWs and detection]
{Triplets of supermassive black holes:\\ Astrophysics, Gravitational Waves and Detection}

\author[Pau Amaro-Seoane et al.]{Pau Amaro-Seoane$^{1}$\thanks{e-mail:
Pau.Amaro-Seoane@aei.mpg.de (PAS)}, Alberto Sesana$^{2}$, Loren
Hoffman$^{3}$,\newauthor Matthew Benacquista$^{4}$, Christoph Eichhorn$^{5,7}$, Junichiro Makino$^6$ \& Rainer Spurzem$^{7,8,9}$
\\
$^{1}$ Max-Planck Institut f\"ur
Gravitationsphysik (Albert-Einstein-Institut), Am M{\"u}hlenberg 1,
D-14476 Potsdam, Germany and\\
Institut de Ci{\`e}ncies de l'Espai (CSIC-IEEC), Campus UAB, Torre C-5, parells, $2^{na}$ planta, ES-08193
Bellaterra, Barcelona, Spain \\
$^{2}$ Penn State University, 104 Davey Lab, \#113 University Park, PA 16802-6300, USA\\
$^{3}$ Northwestern University, Dearborn Observatory, 2131 Tech
Drive, Evanston, IL 60208-2900, USA\\ 
$^{4}$ {Center for Gravitational Wave
Astronomy, University of Texas at Brownsville, Brownsville, TX 78520,
USA} \\ 
$^{5}$ Institut f\"ur
Raumfahrtsysteme, Universit\"at Stuttgart, Pfaffenwaldring 31, D-70550
Stuttgart, Germany \\ 
$^{6}$ Division of Theoretical Astronomy, National
Astronomical Observatory, 2-21-1 Osawa, Mitaka, Tokyo 181-8588, Japan\\ 
$^{7}$ National Astronomical Observatories of China, Chinese Academy of
Sciences, 20A Datun Lu, Chaoyang District, 100012, Beijing, China\\
$^{8}$ Kavli Institute for Astronomy and Astrophysics, Peking University, China \\
$^{9}$ Astronomisches Rechen-Institut, M{\"o}nchhofstra{\ss}e 12-14, 69120,
Zentrum f\"ur Astronomie, Universit\"at Heidelberg, Germany\\
}

\date{\today, submitted to MNRAS}

\pagerange{\pageref{firstpage}--\pageref{lastpage}} \pubyear{\today, submitted to MNRAS}

\begin{document}

\label{firstpage}

\maketitle

\begin{abstract}
Supermassive black holes (SMBHs) found in the centers of many galaxies
are understood to play a fundamental, active role in the cosmological structure 
formation process. In hierarchical formation scenarios, SMBHs 
are expected to form binaries following the merger of their host galaxies. If these
binaries do not coalesce before the merger with a third galaxy, the formation
of a black hole triple system is possible. Numerical simulations of the
dynamics of triples within galaxy cores exhibit phases of very high
eccentricity (as high as $e \sim 0.99$). During these phases, 
intense bursts of gravitational radiation can be emitted at orbital periapsis, which produces a gravitational wave signal at frequencies substantially higher 
than the orbital frequency. The likelihood of detection of these bursts with 
pulsar timing and the Laser Interferometer Space Antenna ({\it LISA}) is estimated 
using several population models of SMBHs with masses $\gtrsim 10^7~{\rm M_\odot}$. 
Assuming 10\% or more of binaries are in triple systems, 
we find that up to a few dozen of these bursts 
will produce residuals $>1$ ns, within the sensitivity range of forthcoming pulsar 
timing arrays (PTAs). However, most of such bursts will be washed out in the 
underlying confusion noise produced by all the other 'standard' SMBH binaries
emitting in the same frequency window. A detailed data analysis study would be required 
to assess resolvability of such sources. Implementing a basic resolvability
criterion, we find that the chance of catching a resolvable burst
at a one nanosecond precision level is $2-50$\%, depending on the adopted SMBH evolution model. 
On the other hand, the probability of detecting bursts produced by 
massive binaries (masses $\gtrsim 10^7\msun$) with {\it LISA} is negligible.
\end{abstract}


\section{Introduction} 
\label{intro}
It is well established that most galaxies host supermassive black holes 
(SMBHs) in their centers \citep{rich98}. In the past decade, compelling evidence of the
correlation between the mass of the central SMBH and the bulge  velocity
dispersion and luminosity has been collected
~\citep{ferrarese00,gebhardt00,merritt01,tremaine02}, indicating
a coevolutionary scenario for SMBHs and their hosts. On a
cosmological scale, galaxy formation and evolution can be understood by
semi-analytic modeling, where properties of the baryonic matter are followed in
the evolving dark matter halos obtained from large-scale models of hierarchical
gravitational structure formation. A simple model of galaxy and central SMBH
evolution in which every merger of galaxies leads quickly to coalescence of
their central black holes can quantitatively reproduce both the SMBH mass-bulge
luminosity relation~\citep{kauffmann00} and the SMBH mass-velocity dispersion
relation~\citep{haehnelt00}. 

In this general picture, if both of the galaxies involved in a merger 
host a SMBH, then the formation of a SMBH binary is an inevitable 
stage of the merging process. Following the merger, the two black holes sink
to the center of the merger remnant because of dynamical friction ~\citep{begelman80}. 
When the mass (either in gas or stars) enclosed in their orbit is of the 
order of their own mass, they start to feel the gravitational pull of each other, 
forming a bound binary. The subsequent binary evolution is, however, still unclear. 
In order to coalesce, the binary must shed its binding energy and 
angular momentum; a dynamical process known in literature as `hardening'.
A crucial point in assessing the fate of the binary is the efficiency with which it transfers
energy and angular momentum to the surrounding gas and stars. 

The case of SMBH binaries in stellar environments has received a lot of attention 
in the last decade. The system is usually modeled as a massive binary 
embedded in a stellar background with a given phase space distribution. The region of
phase space containing stars that can interact with the SMBH binary in one
orbital period is known as the loss cone~\citep{frankr76,AS01,milos03}. 
As the binary evolves, it ejects stars on intersecting orbits via the 
so called `slingshot mechanism', causing a progressive emptying of the loss cone,
which ultimately increases the hardening time scale. 
Without an efficient physical mechanism for repopulating the loss cone,
the binary will never proceed to small separations where coalescence induced by 
gravitational radiation takes place within a Hubble time. This is known 
as the stalling or `last parsec' problem \citep{milos01}.

In the last decade, several
solutions to the stalling issue have been proposed. Axisymmetric 
or triaxial stellar distributions may significantly shorten the coalescence timescale
\citep{yu02,merritt04,bercziketal06}. This is bacause the presence of deviations from spherical 
symmetry can produce ``boxy'' orbits, as seen
by \cite{bercziketal06}.  These orbits produce centrophilic stellar
orbits and, therefore, replenish the loss-cone.  However, more recent calculations by \cite{AmaroSeoaneSantamaria09} of the outcome
of the merger of two clusters initially in parabolic orbits \citep{ASF06} have not been able to reproduce the rotation necessary to create the unstable bar structure. Other studies have invoked eccentricities
of the binary to refill the loss cone, since this effect could alter the cross
section for super-elastic scatterings (thus altering the state of the loss
cone) and shorten the gap to the onset of gravitational radiation
effects (e.g.:~\citealt{hemsendorf02, aarseth03, bercziketal06,ASF06,ASMF09}).
The presence of massive perturbers may also help replenishing the loss 
cone, boosting the binary hardening rate \citep{perets07}. 
On the other hand, in smooth particle hydrodynamics simulations of SMBH binaries 
in gas-rich environments, efficient hardening induced by the tidal 
interaction between the binary and the gas medium 
has been observed, indicating a possible quick coalescence 
\citep{escala05,dotti06}. However, current simulations do not have the 
resolution to follow  the binary fate down to the gravitational wave (GW) 
emission regime, and robust conclusions about its late inspiral and coalescence can not be drawn.
In any case, very massive low redshift systems, which are the
major focus of our study, are more likely to reside in massive gas poor galaxies
and their dynamics is probably dominated by stellar interactions.

When scaled to very massive binaries (masses $>10^8\msun$), the 
inferred coalescence timescales in a stellar dominated environment
are of the order of few Gyrs, indicating that SMBH binaries may be relatively long living systems. 
If the typical timescale between two subsequent mergers is comparable the SMBH
binary lifetime, then a third black hole may reach the nucleus when the binary is still
in place, and the formation of SMBH triplets might be a common step in the galaxy
formation process. Recent studies of 
galaxy pairs lead to the conclusion that $30-70$\% of present day massive galaxies
have undergone a major merger since redshift one \citep{bell06,lin08}, where 'major' 
means with baryonic mass ratio of the two components larger than $1/3$ or $1/4$ (depending on the study),
which is a quite conservative threshold. This means that, on average, all massive galaxies
have experienced a merger event in the last ten billion years. Assuming uncorrelated
events, and a typical binary lifetime of one billion years, then 10\% of 
SMBH binaries may form a triplet. With increasing redshift (and decreasing masses), dynamical 
timescales become shorter and shorter, implying that triplets may have been more 
common in the high redshift Universe. 

In this paper we focus on SMBH triplets, studying their dynamical evolution, 
GW emission, and detectability. 
Employing sophisticated three body scattering experiments calibrated on 
direct-summation {\sc Nbody} simulations, we study the dynamical evolution 
of the system, finding surprisingly high eccentricities of the inner SMBH 
binary (up to $e>0.99$). Even though the triple interaction would possibly lead 
to an ejection of one or even all SMBHs \citep{valtonen94}, most of the
systems are long living ($\sim10^9$ yrs, \cite{HoffmanLoeb07}), and 
final coalescence is more common than ejection, confirming 
analytical results by \cite{makino94}. We model at the leading quadrupole 
order \citep{peters63} the bursts of gravitational radiation emitted in the highly eccentric 
phase, assessing detectability with future GW experiments. 
Adopting cosmologically and astrophysically 
motivated models for SMBH formation and evolution, we estimate
reliable event rates.

In order to cover the low frequencies generated by the 
expected cosmological population of coalescing SMBH binaries 
\citep[e.g.,][]{wl03,ses04,ses05,ses07} or plunges of compact
objects such as stellar black holes on to supermassive ones \citep[see
e.g.][for a review and references therein]{Amaro-SeoaneEtAl07}, the space-born
observatory {{\it LISA}} \citep{bender98} has been planned to be covering the range of frequencies of
$\sim10^{-4}-10^{-1}~{\rm Hz}$. Moving to even lower frequencies, the Parkes Pulsar
Timing Array \citep[PPTA,~][]{manchester06,manchester07}, the European Pulsar
Timing Array \citep[EPTA,~][]{EPTA} and the North American Nanohertz
Observatory for Gravitational Waves \citep[NANOGrav,~][]{NANOGrav} are already
collecting data and improving their sensitivity in the frequency range of
$\sim10^{-8}-10^{-6}$ Hz, and in the next decade the planned Square Kilometer Array 
\citep[SKA,~][]{laz09} will provide a major leap in sensitivity. 

Throughout this paper we consider only very massive systems, with total mass $\sim 10^8\msun$.
Our goal is to investigate if the high frequency nature of eccentric bursts
can provide information about systems which would otherwise emit outside the
frequency windows of the planned GW experiments quoted above, by shifting 
wide (separation $\gtrsim0.1$ pc) SMBH binaries into the PTA window or by boosting
relatively massive (masses $>10^7\msun$) systems into the {\it LISA} domain. We note that the bursts analyzed here are different from the `bursts with memory', which arise during the actual coalescence of SMBH binaries and are discussed in~\citet{pshirkov09} and~\citet{vanhaasteren09}.

The structure of the paper is as follows. In Section \ref{astro}, we describe our
comprehensive study of the dynamics of triple systems and investigate the
eccentricity evolution of the inner binary by using direct-summation $N-$body
techniques and a statistical 3-body sample calibrated on the $N-$body results.
In Section \ref{gravwaves}, we model the GW signal produced by eccentric 
bursts and we introduce observable quantities for PTAs and {{\it LISA}}.
In Section 4 we construct detailed populations of emitting SMBH binaries and 
triplets, and we discuss our results in terms of signal
observability and detection rates in Section 5. Lastly, we briefly summarize 
our results in Section 6.  

\section{Dynamics of triple systems} \label{astro}

In modeling the dynamics of black hole triple systems within the centres of
galaxy merger remnants, direct $N$-body integrations provide the most accuracy
but are the most computationally expensive. We performed eight direct $N$-body
calculations and used these to test the validity of an approximation scheme
involving three-body SMBH dynamics embedded in a smoothed galactic potential
with dynamical friction and gravitational radiation modeled by drag forces.

\subsection{Direct $N-$body calculations}\label{nbody}

The direct-summation {\sc Nbody} method we employed for all the
calculations includes the {\em KS regularisation}. Thus, when
two particles are tightly bound to each other or the separation between
them becomes very small during a hyperbolic encounter, the system
becomes a candidate to be regularised in order to avoid problematical
small individual time steps \citep{KS1965}. This procedure was later
exported to systems involving more than two particles. In particular,
the {\em KS regularisation} has been adapted to isolated and perturbed
3-- and 4--body systems---the so-called {\em triple} (unperturbed
3-body subsystems), {\em quad} (unperturbed 4-body subsystems) and the
{\em chain regularisation}. The latter is invoked in our simulations
whenever a regularised pair has a close encounter with another single
star or another pair \citep{aarseth03b}.

The basis of direct {\sc Nbody} codes relies on an improved {Hermite}
integrator scheme \citep{Aarseth1999} for which we need not only the
accelerations but also their time derivative. The computational effort
translates into accuracy so that we can reliably keep track of
the orbital evolution of every particle in our system.  
In order to make a highly accurate estimate of the eccentricity evolution 
of the SMBH system, we do not employ a softening to the gravitational force (i.e.
substituting the $1/r^2$ factor with $1/(r^2+\epsilon^2)$, where $r$ is
the {separation} and $\epsilon$ the softening parameter) that weakens 
the interaction at small separations.

\begin{figure}
\resizebox{\hsize}{!}{\includegraphics[scale=1,clip]{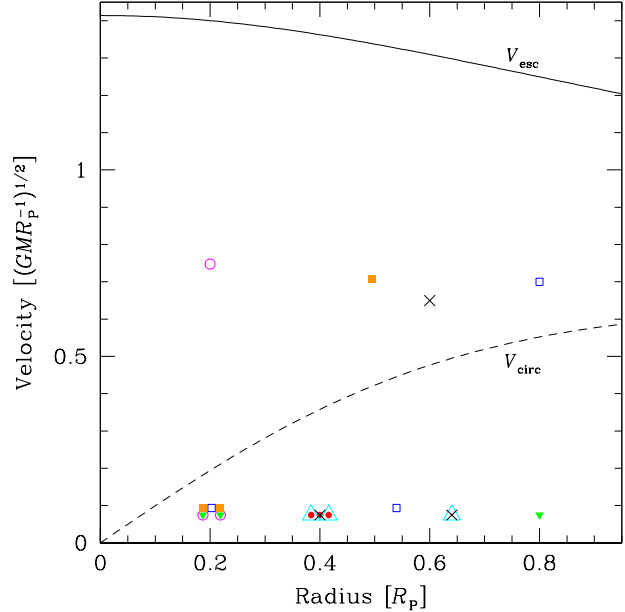}} 
\caption{
Initial conditions for the different direct $N$-body simulations in the $R_{\rm
P}$, $V$ plane. For each simulation we choose the separation between two SMBHs
to be substantially smaller than the distance to the third SMBH. The initial
parameters are selected in such a way that they are random but with an initial
velocity smaller than the escape velocity $V_{\rm esc}$ and with a radius $r < R_{\rm P}$.
We also show the circular velocity in the figure, $V_{\rm circ}$.
Initially The set of 3 SMBHs of simulation A is represented with red solid
bullets; for simulation B with green solid triangles; for simulation C with
cyan open triangles; for simulation D with blue open squares; for simulation E
with pink open circles; for simulation H with solid orange squares and for
simulations F and G with black crosses. In the case of simulations A, B, C, D,
E, F and G the three SMBHs are set initially in a planar configuration. In the
case of simulation H they have a z component different from zero in both the
coordinates and velocities.  In the cases of simulations A, B, C, E and H we
slightly modified the positions of the symbols in order to avoid an overlap
\label{fig.init_cond_SM} } 
\end{figure}

\begin{table*}  
\begin{center}
\begin{tabular}{l|*{7}{c|}c} \hline 
Model &  A & B & C & D & E & F & G & H\\ 
\hline 
${\cal N}_{\star}$ &  64,000 & 64,000 & 64,000 &  64,000 & 64,000  &
64,000  & 512,000 & 256,000 \\ 
\hline

$R/{R}_{\rm P}$ & 0.40012 & 0.80006 & 0.40012 & 0.80006 & 0.20025 &
0.60017 & 0.60017 & 0.49497\\
& 0.40012 & 0.20303 & 0.64031 & 0.53956 & 0.20303 & 0.64031 & 0.64031 &
0.20278\\ 
& 0.40012 & 0.20303 & 0.40012 & 0.20303 & 0.20303 & 0.40012 & 0.40012 &
0.20278\\ 
\hline

$V/{V}_{\rm esc}$ & 0.07476 & 0.07476 & 0.07476 & 0.70000 & 0.74762 &
0.64991 & 0.64991 & 0.70711\\
& 0.07476 & 0.07476 & 0.07476 & 0.09345 & 0.07476 & 0.07476 & 0.07476 &
0.09345\\
& 0.07476 & 0.07476 & 0.07476 & 0.09345 & 0.07476 & 0.07476 & 0.07476 &
0.09345\\ 
\hline 
$V/{V}_{\rm circ}$
& 0.20886 & 0.13543 & 0.20886 & 1.26802 & 3.84516 & 1.36391 & 1.36391 &
1.68376\\ 
& 0.20886 & 0.37956 & 0.15108 & 0.20886 & 0.37956 & 0.15108 & 0.15108 &
0.47496\\ 
& 0.20886 & 0.37956 & 0.20886 & 0.47442 & 0.37956 & 0.20886 & 0.20886 &
0.47496\\ 
\hline 
\end{tabular} 
\end{center} 
\label{tab.init_cond}
\caption{Initial conditions for the set of three SMBHs in each of the
{eight} direct $N$-body simulations. ${\cal N}_{\star}$ is the total number of stars
employed in the simulation, ${V}$ their velocity, $R$ their position,
${R}_{\rm P}$ the position of the SMBHs in terms of the Plummer
radius, ${V}_{\rm esc}$ their escape velocity and ${V}_{\rm circ}$ is
their circular velocity. The mass of the SMBHs in $N-$body units is $1\cdot10^{-2}$,
they are equal-mass and the mass of a star $1.15\cdot10^{-5}$}
\end{table*}

The initial conditions for the set of three SMBHs, used to conduct an
exploration of the initial parameter space are shown in
Table~\ref{tab.init_cond}. For the stellar system, we use a Plummer model~\citep{Plummer11}, which is an $n=5$ polytrope with a compact core and an extended outer envelope.  In this model the
density is approximately constant in the centre and drops to zero in the outskirts, $\phi=-{GM_{\star}}/{\sqrt{r^2+R_{\rm P}^2}}$, with $M_{\star}$ the
total stellar mass. This defines the Plummer radius  $R_{\rm P}$. We
depict the initial conditions in Figure~\ref{fig.init_cond_SM}, relative to the
circular and escape velocity of the Plummer potential.  We present results from
eight direct numerical simulations, one using 512,000 stars using the
special-purpose GRAPE6 system and the remaining simulations using Beowulf PC
clusters and the AEI mini-PCI GRAPE cluster {\sc Tuffstein}.

\subsection{Three-body improved statistics}\label{3body}

While direct $N$-body simulations yield a very accurate result, they
should be seen as a way to calibrate and test faster, more approximate
simulations which can exhaustively cover the parameter space and provide good
statistics.  We note that the SMBHs in the $N-$body simulations
are equal-mass and (with the exception of simulation H), all of the systems studied 
with this method are coplanar. This was done because setting all SMBHs on a single 
plane accelerates the dynamics, shortening the integration time. 

In general, one wants to explore the whole parameter space, including non 
coplanar systems with different SMBH masses.  For
this purpose, we performed an ensemble of 1000 three-body experiments, with the
three Euler angles of the outer orbit sampled uniformly and a distribution of
mass ratios motivated by Extended Press-Schechter theory (with typical mass
ratios ${\rm m_{1}:m_{2}:m_{3}}$ around 3.5:1).  In each experiment we computed
the Newtonian orbits of three SMBHs embedded in a smooth galactic potential and
added drag forces to account for gravitational radiation and dynamical
friction.  We also included coalescence conditions when either of the two SMBHs pass within three Schwarzschild radii of each other, or the gravitational radiation timescale
becomes short relative to the orbital period of the binary.
Close triple encounters were treated using a KS-regularised few-body
code provided by Sverre Aarseth \citep{MA90,MA93}, while the two-body motion in
between close encounters was followed with a simple 4th-order Runge-Kutta
integrator.  See \cite{HoffmanLoeb07} for further details on the code.  The
initial conditions are those for the {\em canonical} ICs as in
\cite{HoffmanLoeb07}. We performed each run twice---once with gravitational radiation drag and the coalescence conditions, and once without.

The 3-body experiments are divided into two computational regimes based upon a dimensionless parameter, $\alpha$, that measures the relative tidal perturbation to the inner binary by the interloper at apoapsis:
\begin{equation}
\alpha = 2 \frac{R_{\rm apo}^3\,M_{\rm single}}
                    {M_{\rm bin,\,smaller}\,D_{\rm single}^3},
\end{equation}
where $R_{\rm apo}$ is the apoapsis separation of the two inner binary members,
$M_{\rm single}$ is mass of the single SMBH, $M_{\rm bin,\, smaller}$ is mass
of the smaller inner binary member and $D_{\rm single}$ is the distance of the
interloper (single SMBH) from the inner binary centre-of-mass. In the limit $\alpha \rightarrow$ 0, we know that the period of the inner binary is perfectly
Keplerian (plus gravitational radiation), since the perturbation to the force from the third body is
negligible, and thus we can do orbit-averaged integration instead of precisely integrating
the trajectories of all three bodies. The two regimes are defined as follows:
\begin{enumerate}
\item {\em The first regime} corresponds to 
when a 3-body interaction is taking place (defined by $\alpha > 10^{-5}$), an extremely conservative
criterion for when we need to do the full three-body integration.

\item {\em The second regime} corresponds to when the single SMBH and binary are wandering separately through the
galaxy ($\alpha <10^{-5}$), often on the order of a Hubble time.
\end{enumerate}
In regime (i) the
3-SMBH dynamics is integrated using Sverre Aarseth's high-precision, regularised
CHAIN code.  Gravitational radiation and stellar-dynamical friction are treated
as perturbing, velocity-dependent forces on the three separate bodies.  In
regime (ii) the separate orbits of the single and binary centre-of-mass are followed using
a simple 4th-order Runge-Kutta integrator and the evolution of the binary
semi-major axis and eccentricity are evolved using orbit-averaged equations.

\begin{figure*}
\resizebox{1.0\hsize}{!}{\includegraphics[scale=1,clip=true]{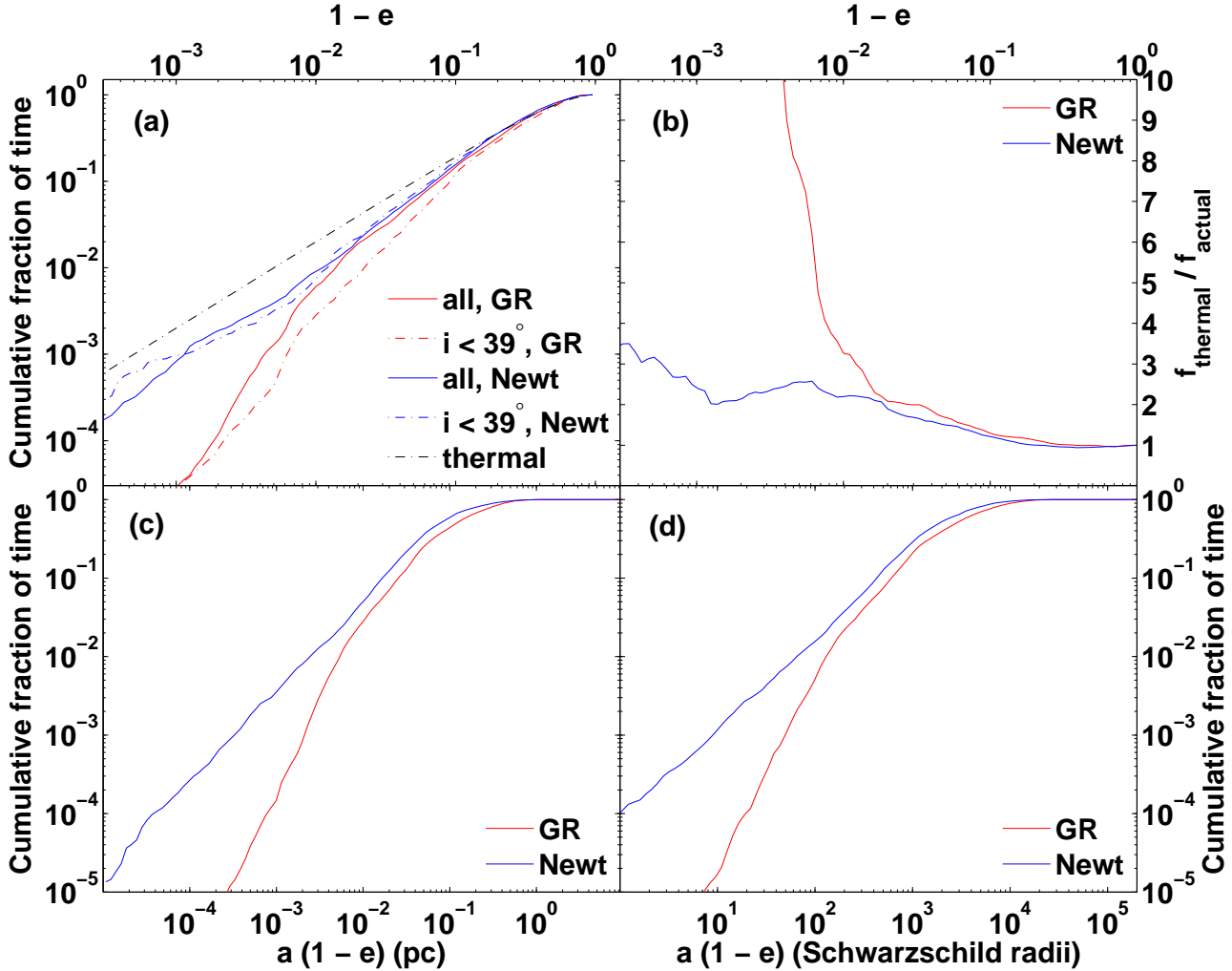}}
\caption{
Cumulative fraction of time for the set of 1000 three-body simulations.
The red, solid line corresponds to all simulations with the 2.5
post-Newtonian correction term; the blue, solid line corresponds to the
same simulations but without it (i.e. purely Newtonian); the dot-dashed
red curve is like the first case but taking into account only systems in
which the third SMBH had an inclination below the critical Kozai angle
of $39^{\circ}$, to compare with the direct $N$-body simulations of
Fig.(\ref{fig.CumulativeFraction_Nbody}); the dot-dashed blue curve is
the same, but for the Newtonian cases and the dot-dashed black curve
corresponds to the thermal distribution, since the direct $N$-body
simulations do not have relativistic correction terms. The top left
panel shows the eccentricity computed from the instantaneous positions
and velocities of the two binary members, and the top right panel shows
the ratio of the actual to the thermal distribution.  For the Newtonian
runs, the distribution is within a factor of 2 of thermal down to
$1-e=0.01$ and within a factor of 3 of thermal down to  $\sim
1-e=0.001$.  The bottom two panels show the pericenter distances of the
binary in pc and in units of the sum of the Schwarzschild radii of the
two binary members.  No coalescence was allowed during close encounters
for the Newtonian runs.  We can see that the distributions
converge, with respect to the statistics, from the fact that they are
substantially the same as the lower-number experiments and that the
Newtonian and gravitational radiation distributions match at low eccentricities
\label{fig.eccFigure_3bSimulations}
}
\end{figure*}

\begin{figure*}
\resizebox{\hsize}{!}{\includegraphics[scale=1,clip]{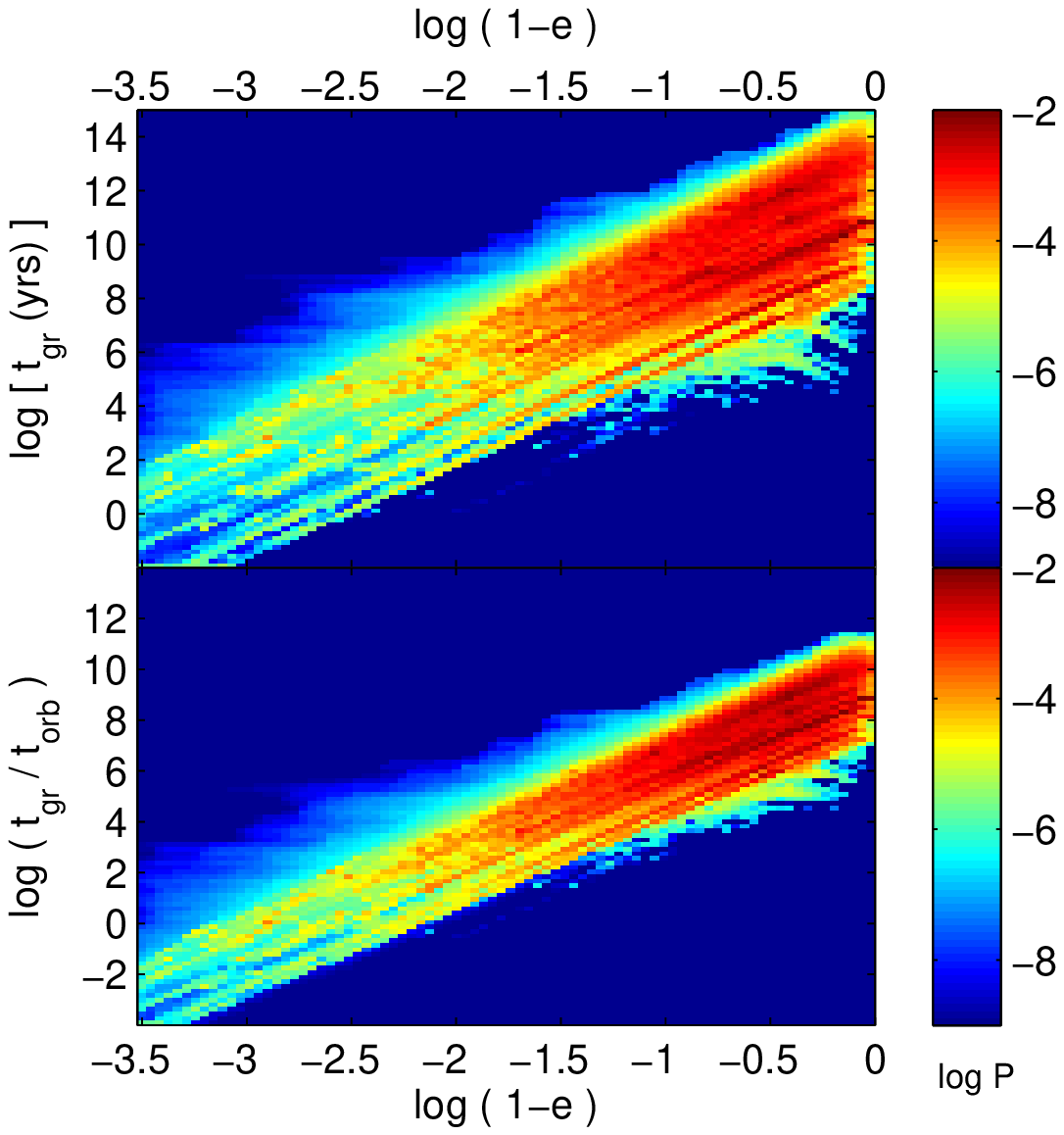}}
\caption{
Time spent at different locations in the $t_{\rm gr}$ - (1-e) plane in the Newtonian simulations
(blue lines) of Figure 2.  $P = t/t_{\rm close}$ is the probability of finding the binary in a given bin
at a randomly chosen time, where $t_{\rm close}$ is the total simulation time spent in close 3-body
encounters. The simulation being presented here is the same as the Newtonian simulation (blue lines) 
in Figure 2. The purpose of this figure is to understand whether gravitational radiation is the
main reason for the fall-off in the eccentricity distribution, since the highest eccentricity systems
have short coalescence times and quickly disappear.  $t_{\rm gr}$ is in years in the upper panel and
in orbital periods in the lower panel (the typical resonant encounter takes on order $10^5$ yrs).
Note that 1-e = 0.002 is about where $t_{\rm gr}$ falls to less than an orbital period
\label{fig.gravRadn}
}
\end{figure*}

In computational regime (ii) (between 3-body encounters), the stellar interactions are treated using the hard binary prescription of~\citet{quinlan96}. The eccentricity evolution of the inner binary under stellar interactions for near equal-mass hard binaries is quite weak so it is neglected entirely and only the binary semi-major axis is evolved under stellar interactions. The eccentricity is evolved under gravitational radiation as given by~\citet{peters64}. Dynamical friction tends to increase the eccentricity of a binary in the supersonic regime (where the orbital speed exceeds the stellar velocity dispersion), and to circularize it in the subsonic regime. Since the triple SMBH system starts out supersonic in these 3-body experiments, this effect produces a slight increase in the outer binary eccentricity during the initial inspiral of the third SMBH. Although we neglect the eccentricity evolution of the inner binary during this phase, we find that its eccentricity is thermalized by the first resonant 3-body encounter. The timescale of the chaotic 3-body interactions ($\sim 10^5$ yr) is much shorter than the stellar-dynamical timescale ($~ 10^7-10^{10}$ yr), so any effect of stellar interactions during these encounters is completely negligible. 
Thus, stellar interactions play only two roles in our 3-body simulations:
\begin{enumerate}\renewcommand{\labelenumi}{\alph{enumi}}
\item They bring the third BH in to interact with the inner binary from an
initial marginally stable hierarchical triple configuration;
\item During phase (ii), stellar-dynamical interactions gradually decrease the
binary semi-major axis, so that it enters the next three-body encounter harder
than it left the last one if the time between encounters is long ( $> ~10^7$
yrs).  The binary may even coalesce between encounters due to this gradual
shrinking.
\end{enumerate}

Consequently, the distribution in eccentricities is best estimated using the fraction of time that binaries spend at a given eccentricity while in computational regime (i) since there is minimal evolution of the eccentricity while in regime (ii). We note that the overwhelming majority of the time spent in regime (i) is still spent with $\alpha$ small enough that the system can be though of as a separate inner and outer binary, and so the instantaneous inner binary semi-major axis and eccentricity are well-defined.

\subsection{Distribution of Eccentricities}

Figure~\ref{fig.eccFigure_3bSimulations}a shows the fraction of the time
that the binary (closest SMBH pair) spends above a given eccentricity
during close encounters, averaged over 1000 three-body experiments.  The red solid curves  are
the results of the standard runs including gravitational radiation drag and coalescence conditions 
while the blue solid curves are the ``Newtonian'' case with these effects neglected.
The dashed red and blue curves are averaged over only
those experiments where the initial BH configuration has inclination
$<39^{\circ}$, the critical angle for Kozai oscillations, for comparison with
the direct $N$-body simulations that use coplanar initial conditions.  The black (dot-dashed) line shows the thermal distribution of eccentricities
for reference purposes.  Three-body interactions result
in a thermal distribution of eccentricities, truncated at very high
eccentricities by coalescence in collisions when gravitational radiation is included.  The
similarity of the dashed and solid curves shows that once the initial secular
evolution is over, the system quickly thermalizes and little memory of the
initial configuration is maintained. Figure~\ref{fig.eccFigure_3bSimulations}b
shows the ratio of the {thermal} to the {actual} distribution as a function of
eccentricity. The first close encounter in each experiment has
been excluded from these plots, since it begins from a stable hierarchical triple
configuration and includes a long period of secular evolution, whereas chaotic
three-body encounters are the focus of this work. 

\begin{figure*}
\resizebox{\hsize}{!}{\includegraphics[scale=1,clip]{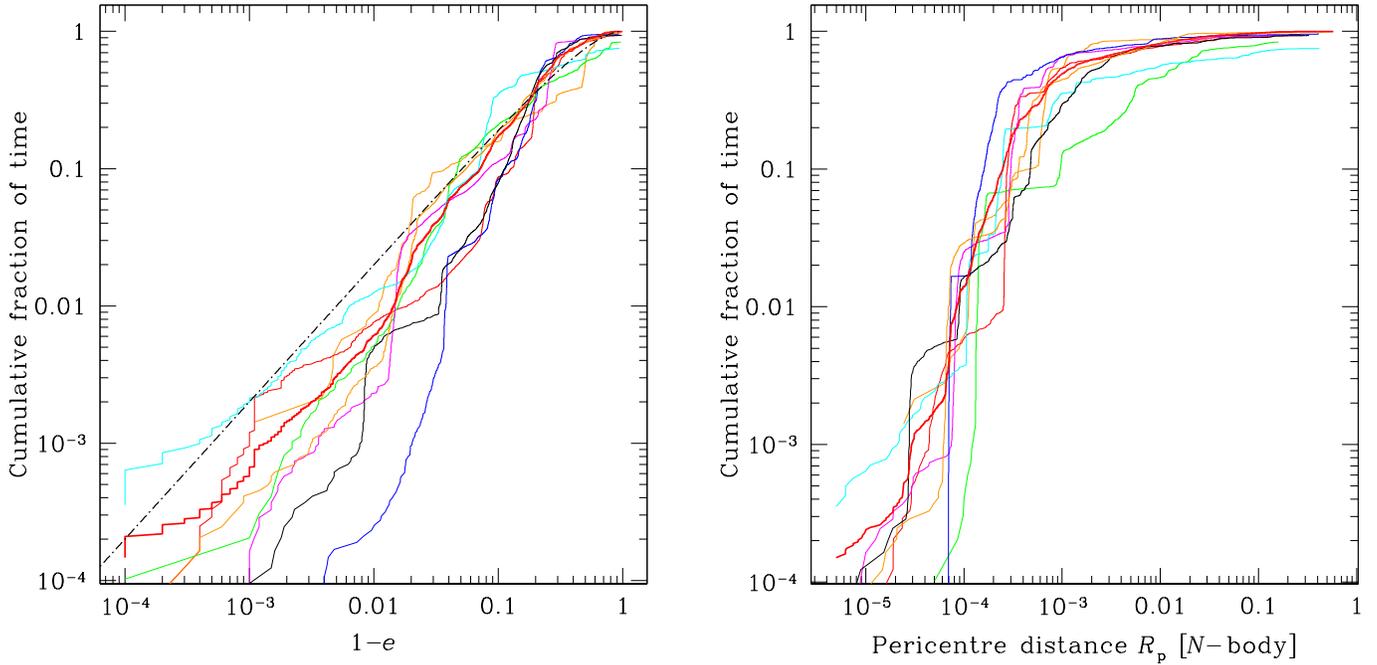}}
\caption{ Left panel: Fraction of time for the direct $N$-body
simulations at which the binary black hole has eccentricity in the range
$1-e$ for the 512,000 stars simulation (thin green curve) and for the
lower-resolution simulations. The thick red curve corresponds to the
average of these lower-resolution computations and the dashed curve to
the thermal distribution. Right panel: Cumulative fraction of time spent
on a certain periapsis distance for all direct $N$-body simulations
following the same colour labelling. One $N-$body unit of distance is
${\cal U}|_{\rm R}=1.1$ pc
\label{fig.CumulativeFraction_Nbody} }
\end{figure*}

The runs with and without gravitational radiation closely follow each other and
the thermal distribution up to eccentricities $e \sim 0.99$.  At higher
eccentricities, the Newtonian distribution remains within a factor of 2 to 3 of
the thermal distribution, but the gravitational radiation curve falls off
sharply, since these high-eccentricity systems coalesce quickly through
emission of gravitational waves.  To verify this interpretation of
Figure~\ref{fig.eccFigure_3bSimulations}, we plot the time spent at different
locations in the $t_{gr}$-$(1-e)$ plane in Figure~\ref{fig.gravRadn}, where the
gravitational radiation timescale is computed from the instantaneous $(a,e)$ of the
binary.  The eccentricity where the red and blue curves
diverge in Figure~\ref{fig.eccFigure_3bSimulations}b ($1-e \sim 0.01$) is the value where
the typical $t_{gr}$ falls to just a few orbital periods, so that the
binary can coalesce quickly by gravitational radiation before the third
body scatters it on to a lower-eccentricity orbit.
Figures~\ref{fig.eccFigure_3bSimulations}c-d show the time spent by the
binary at various pericenter separations, in pc and in Schwarzschild
radii.  Note that the gravitational radiation curve diverges sharply
from the Newtonian one when the pericenter separation reaches $\sim$100
Schwarzschild radii.
We show the fraction of time spent at different eccentricities and the fraction
of time spent at different pericenter separations for the $N-$body simulations
in Figure~\ref{fig.CumulativeFraction_Nbody}.  The qualitative features are
retained.

\section{Gravitational waves: analysis of the signal} 
\label{gravwaves}

In this section we make use of the leading Newtonian order derivation
of the GW radiation from eccentric binaries, as described 
in~\citet{peters63}. We also use of geometric units, with  $G=c=1$.
Consider a system with masses $M_2<M_1$ orbiting with an orbital 
{\it rest frame} frequency
$f_r=\omega/2\pi$, and with eccentricity $e$; at the quadrupole leading order, 
the luminosity $\dot{E}$ emitted by the system {\it averaged over one complete 
orbit} is:

\begin{equation}
\dot{E}=\frac{32}{5}{\cal M}^{10/3}(2\pi f_r)^{10/3}F(e)=\dot{E}_cF(e) 
\label{luminosity}
\end{equation}
where 
\begin{equation}
F(e)=\sum_{n=0}^\infty g(n,e)= 
\frac{1+\frac{73}{24}e^{2}+\frac{37}{96}e^{4}}{(1-e^{2})^{7/2}},
\label{Fe}
\end{equation}
and ${\cal M}=M_1^{3/5}M_2^{3/5}/(M_1+M_2)^{1/5}$ is the chirp mass of the system. 
$\dot{E}_c$, defined by the right-hand side in equation (\ref{luminosity}), is the luminosity
emitted by a circular binary orbiting at the same frequency $f_r$. 
The binary radiates GWs in the whole spectrum of harmonics 
$f_{r,n}=nf_r\,\,\,(n=1, 2, ...)$, and the relative power radiated in each 
single harmonic is described by the function $g(n,e)$, defined as:

\begin{equation} 
g(n,e) = \frac{n^4}{32}\left(B_n^2 + \left(1-e^2\right)A_n^2 +
\frac{4}{3n^2}J_n(ne)^2\right) ,
\end{equation}
where $J_n(x)$ are the Bessel functions and $A_n$ and $B_n$ are also
defined in terms of the $J_n$ as:
\begin{eqnarray} 
B_n & = & J_{n-2}(ne)-2eJ_{n-1}(ne) + \frac{2}{n} J_n(ne) \nonumber \\ 
& & +2eJ_{n+1}(ne)  - J_{n+2}(ne) \\ 
A_n & = & J_{n-2}(ne) - 2J_n(ne) + J_{n+2}(ne).
\end{eqnarray}
The total luminosity of the source can then be written, using equations (\ref{luminosity}) and (\ref{Fe}), as the sum 
of the component radiated at each single harmonic:
\begin{equation}
\dot{E}=\sum_{n=0}^\infty \dot{E}_n=\sum_{n=0}^\infty \dot{E}_c g(n,e). 
\end{equation}

Given a general GW characterised by the two polarised component waves
$h_+$ and $h_\times$, the rms amplitude of the wave is defined
as $h=\sqrt{\langle h_+^2 + h_\times^2\rangle}$, where $\langle\ \ \rangle$
denotes the average over directions and over time. The flux radiated 
in the GW field is related to the derivatives of its amplitude components
by the relation~\citep{thorne87} 
\begin{equation}
\frac{dE}{dtdA}=\frac{1}{16\pi}\left(\dot{h}_+^2 + \dot{h}_\times^2\right). 
\label{eflux}
\end{equation}
The sinusoidal nature of the waves implies 
$\langle \dot{h}_+^2 + 
\dot{h}_\times^2\rangle=4\pi^2f_r^2\langle h_+^2 + h_\times^2\rangle$. 
So that integrating equation (\ref{eflux})
over a spherical surface of radius $d_L$ (the luminosity distance
from the source) centered at the source  
and averaging over an orbital period, directly 
relates $\dot{E}$ to the wave rms amplitude. 
We can then infer that the rms amplitude and the energy radiated in the
n-th harmonic are related as~\citep{finn00}
\begin{equation} 
h_n=\frac{1}{\pi d}\frac{\sqrt{\dot{E}_n}}{f_{r,n}}=2\sqrt{\frac{32}{5}}\frac{{\cal M}^{5/3}}{nd_L}(2\pi f_r)^{2/3}\sqrt{g(n,e)},
\label{hrms}
\end{equation}
where $d=d_L/(1+z)$.
In the limit of a circular orbit (i.e., $g(n,e)=\delta_{n,2}$ in the
Kronecker-$\delta$ notation), equation~(\ref{hrms}) returns the usual sky-polarization averaged
amplitude~\citep{thorne87}. 

\subsection{Observed quantities}\label{obsquants}
Since we are interested in an estimate
of the detectability of extremely eccentric binaries 
(induced by triple interactions) by means
of pulsar timing (and possibly {{\it LISA}}) observations, we first 
introduce an extension of the {\em characteristic amplitude} 
to include eccentric binaries. Eccentric binaries emit 
pulses of GWs at their periapsis passages, and the rms amplitude 
of each harmonic is given by equation (\ref{hrms}). However, 
$h_n$ is an average amplitude related to the average luminosity
along the orbit. The actual relevant time for the burst
is the periapsis passage timescale $T_p=(1-e)^{3/2}T_{\rm orb}$
($T_{\rm orb}$ is the binary orbital period), and 
if the burst is detected, almost all the energy radiated along
the whole orbit is seen on the timescale $T_p$. This means that 
the relevant detectable amplitude of each harmonic during the burst is
\begin{equation} 
h_{\rm obs,n}=h_n\sqrt{{\cal T} f_{r,n}}, 
\label{hobs}
\end{equation}
where the factor ${\cal T}={\rm max}(T_{\rm orb},T_{\rm obs})$
takes into account the fact that, if $T_{\rm orb}<T_{\rm obs}$, 
multiple bursts are visible during the observation.
Equation (\ref{hobs}) is a crude approximation, nevertheless it catches
the basic features of the observed signal: this is given by the rms amplitude
of each single n-th harmonic multiplied by the square root of the cycles
completed by the harmonic in an orbital period, assuming that 
the binary orbit is a fixed ellipse and GW emission does not change the 
orbital parameters. 

The search for GWs using pulsar 
timing data exploits the effect of gravitational 
radiation on the propagation of the radio waves from one (or more) pulsar(s). 
A passing GW would imprint a characteristic signature on the time of 
arrival of radio pulses (e.g.~\citealt{sazhin78,detweiler79,bertotti83}), producing a so called {\em timing residual}.
We refer the reader to~\citet{jenet04} and~\citet{sesana09}, (hereinafter SVV09) for a detailed 
mathematical description of the GW induced residuals.
The residuals
are defined as integrals of the GW during the observation time. For a 
collection of harmonics, the residuals are given by:
\begin{equation} 
R(T)= \int_0^T{\sum_{n=0}^\infty \left(\frac{\alpha^2-\beta^2}{2(1+\gamma)} h_{+,n} +\frac{\alpha\beta}{(1+\gamma)} h_{\times,n}\right)dt},
\label{restd}
\end{equation}
where $\alpha$, $\beta$, and $\gamma$ are the direction
cosines of the pulsar relative to a Cartesian coordinate system defined
with the $z$-axis along the direction of propagation of the
gravitational wave and the $x$ and $y$ axes defining the $+$
polarization. The harmonics of the two polarizations, $h_{+,n}$ and $h_{\times,n}$, can be found in 
Section 3.2 of~\citet{pierro01}.  The rms residual
$\delta t_\mathrm{gw}$ is then formally defined as 
$\sqrt{\langle R(T)^2\rangle}$. 

A simple derivation of the {\it average} timing residual 
$\delta t_\mathrm{gw}$ generated by a circular binary is given by SVV09. 
With the notations adopted above, their equation
(20) reads:
\begin{equation} 
\delta t_\mathrm{gw}(f) = \sqrt{\frac{8}{15}}\frac{h_2}{2\pi f_r}\sqrt{f T_{\rm obs}}\,,
\label{e:deltatgw}
\end{equation}
where the {\it observed} frequency $f$ is related to $f_r$ as $f=f_r/(1+z)$
(being $z$ the redshift of the source), 
the factor $\sqrt{f T_{\rm obs}}$ takes into account for the signal 'build-up' with
the square root of the number of cycles, and $\sqrt{8/15}$ comes from the angle
average of the amplitude of the signal (cf. equation (17)-(21) of SVV09). 
We can generalise this derivation to the case of bursts produced by eccentric 
binaries, relating the $h_{\rm obs,n}$ of each harmonic to the induced residual residual
at its peculiar frequency via:
\begin{equation} 
\delta t_\mathrm{gw}(f_{n}) = \sqrt{\frac{8}{15}}\frac{h_{\rm obs,n}}{2\pi f_{r,n}},
\label{e:deltatgws}
\end{equation}
The total residual can then be assumed to be of the order:
\begin{equation} 
\delta t_\mathrm{gw} = \left(\sum_{n=0}^\infty\delta t_\mathrm{gw}^2(f_{n})\right)^{1/2}.
\label{e:deltatgw2}
\end{equation}
The estimation given in equations 
(\ref{e:deltatgws}) and (\ref{e:deltatgw2}) is justified because the
integral in equation~(\ref{restd}) gives products of sines and cosines of different harmonics, 
that drop to zero when averaged over the observation, leaving only a sum 
of the square signals produced by each single harmonic (those terms 
including ${\rm cos}^2(2\pi f_n t)$ and ${\rm sin}^2(2\pi f_n t)$).
We shall plot, in Section 4, $R(T)$ for selected eccentric bursts, 
and we will see that $\delta t_\mathrm{gw}$ as defined by equations 
(\ref{e:deltatgws}) and (\ref{e:deltatgw2}) gives a good estimate of 
the amplitude of the induced residual.

For inferring {{\it LISA}} detectability, given 
$h_{\rm obs,n}$, an estimate of the signal to noise ratio (SNR) 
in the {{\it LISA}} detector is straightforwardly computed as: 
\begin{equation} 
{\rm SNR}^2=4\sum_{n=0}^\infty \frac{h_{\rm obs,n}^2}{5fS_{f}}, 
\label{snr}
\end{equation}

\noindent
where $S_{f}$ is the one-side noise spectral density of the detector.  We
adopted the $S_{f}$ given in equation (48) of~\citet{barack04}, based on the
{{\it LISA}} Pre-Phase A Report. We extended the sensitivity down to
$10^{-5}$Hz and we considered detection with two independent TDI
interferometers (which implies a gain of a factor of two in $S_{f}$).  The SNR
computed in this way may seem a poor approximation. However, we have checked
the SNRs against those obtained following the procedure given in Section V-B
of~\citet{barack04}, where the binary is consistently evolved with orbit
averaged post-Newtonian equations, and found agreement at a 20-30\% level,
which is acceptable since we are interested in a {\it preliminary estimation}
of source detectability \footnote{The difference is mainly due to the fact that
the orbital parameters change during the strong GW emission burst, and this
is not taken into account in equations (\ref{hobs}) and (\ref{snr}).}. 
 
\subsection{Some heuristic considerations} 

\noindent
The previous derivation can be use to achieve a heuristic understanding of what
we may expect to actually detect.  Let us consider two binaries `1' and `2'
with the same masses, and semimajor axes related as $a_2=a_1(1-e)$ (suppose `2'
is in circular orbit and `1' on a very eccentric orbit, i.e. $1-e\ll1$).
Equations (\ref{luminosity}) and (\ref{Fe}) provide the luminosity {\it
averaged} over an orbital period.  The eccentric binary `1' has an orbital
period $T_1\propto a_1^{3/2}$ .  But, it emits GWs in a short burst of duration
of the order of its periapsis passage that is $T_p\propto[a_1(1-e)]^{3/2}$. The
mean luminosity of the eccentric binary {\it during the periapsis burst} is
then

\begin{equation}
\dot{E}_{1,p}=\dot{E}_1\frac{T_1}{T_p}\propto\frac{F(e)}{a_1^5}(1-e)^{-3/2},
\end{equation} 
where we used the Newtonian relation to switch from $f$ to $a$
in equation (\ref{luminosity}), and we ignored the source 
redshift. According to equations (\ref{eflux}) and (\ref{hrms}), we can write
$h_1\approx\sqrt{\dot{E}_{1,p}}/f_p$, where we make the assumption 
that $f_p$ is the `dominant frequency of the burst', which corresponds 
to the `periapsis frequency', $f_p\sim f_1(1-e)^{-3/2}$. A circular 
binary with semimajor $a_2$ simply emits a periodic wave with amplitude
$h_2\approx\sqrt{\dot{E}_{2}}/f_2$, where $\dot{E}_{2}\propto 1/a_2^5$. 
Remembering that $a_2=a_1(1-e)$ 
and, consequently, $f_p\approx f$, the $h_1/h_2$ ratio reads:
\begin{equation}
\frac{h_1}{h_2}\sim\left[\frac{1+\frac{73}{24}e^{2}+\frac{37}{96}e^{4}}{(1+e)^{7/2}}\right]^{1/2}={\cal O}(1).
\end{equation} 
Since PTAs detect a timing residual that is $\delta t\sim h/f$, 
it follows that $\delta t_1/\delta t_2\sim h_1/h_2$. The timing residual
caused by a burst that happens to be at the right frequency for PTA 
($\sim 10^{-8}$Hz), generated by a very eccentric binary with an orbital 
frequency $f\ll10^{-8}$, is then {\it of the same order of} the residual 
caused by a circular binary emitting at $f=10^{-8}$. The signal 
is, however, quite different and it is spread over a broad frequency band. 
This heuristic consideration suggests that PTA detection of such
extreme events may be rather difficult, because their signal may be
overwhelmed by GW emitted by 'conventional' binaries with shorter periods. 
On the other hand, we might expect some interesting 
effect for {{\it LISA}}, since this mechanism can boost the 
GW frequency by more than three order of magnitudes and signals from 
systems that would emit at much lower frequencies, 
may be shifted into the {{\it LISA}} domain. 

\section{Constructing the signal from binary and triplet population models}

\subsection{Hierarchical models for SMBH evolution}
To draw sensible predictions about the number of expected detectable
GW bursts, we need to model the population of triple systems that form
during the SMBH hierarchical build up. We start by considering the SMBH 
{\it binary}, population. We are mainly interested here in probing 
massive systems $M=M_1+M_2>10^7\msun$, so that we can use catalogs of 
systems extracted from the Millennium Run~\citep{springel05}. 
We employ the very same catalogs used in SVV09; 
the reader is referred to Section 2 of that paper for details,
here we merely summarise the basics of the procedure.
We compile catalogs of galaxy mergers from the semi-analytical model 
of~\citet{bertone07} applied to the Millennium Run. We then associate
a pair of merging SMBHs to each merging pair of spheroids (elliptical galaxies or bulges of
spirals) according to four different SMBH-host 
prescriptions (Section 2.2 of SVV09). Here we consider the three
Tu models presented in SVV09, in which SMBHs correlate with the spheroid
masses according to the relation given by~\citet{tundo07}, and
differ from each other in the adopted accretion prescription:
the Tu-SA model (accretion triggered on to the more massive black hole {\it before} 
the final coalescence), the Tu-DA model (accretion triggered {\it before} the merger
on to both black holes) and the Tu-NA model (accretion triggered {\it after} the 
coalescence). We also investigate the dependence on the adopted SMBH binary 
population by considering the La-SA and Tr-SA models (see SVV09 for details).
The catalogs of coalescing binaries obtained in this way are then properly
weighted over the observable volume shell at each redshift to obtain 
the differential distribution $d^3N/d{\cal M}dzdt_r$, i.e. the 
coalescence rate (the number of coalescences $N$ per unit proper time $dt_r$) 
in the chirp mass and redshift interval $[{\cal M},{\cal M}+d{\cal M}]$ 
and $[z, z+dz]$, respectively. 

\subsection{Signal from SMBH binaries and triplets}
The GW signal can be divided into two contributions---one from the binaries, 
and one from the triplets. We will refer to the latter
as {\it bursting sources}, since we consider the GW bursts they emit
at the periastron in their eccentric phase. In this study, we consider
the binary population emitting in the PTA domain to be composed of
circular systems dynamically driven by GW emission only. The GW signal
is then given by~\citep{sesana08}:
\begin{equation}
h_c^2(f) =\int_0^{\infty} 
dz\int_0^{\infty}d{\cal M}\, \frac{d^3N}{dzd{\cal M} d{\rm ln}f_r}\,
h^2(f_r),
\label{hch2}
\end{equation}
where $h$ is the sky-polarization average of each single source \citep{thorne87}, and 
$d^3N/dzd{\cal M} d{\rm ln}f_r$ is the instantaneous 
population of comoving systems emitting in a given logarithmic 
frequency interval with chirp mass and redshift in the range 
$[{\cal M},{\cal M}+d{\cal M}]$ and $[z, z+dz]$, and is given by:
\begin{equation}
\frac{d^3N}{dzd{\cal M} d{\rm ln}f_r}=(1-{\cal F}_t)\frac{d^3N}{dzd{\cal M} dt_r}\frac{dt_r}{d{\rm ln}f_r},
\label{dndf}
\end{equation} 
where
\begin{equation}
\frac{dt_r}{d{\rm ln}f_r} = \frac{5}{64\pi^{8/3}} {\cal M}^{-5/3}f_r^{-8/3}.
\label{e:fdot}
\end{equation}
In equation (\ref{dndf}), ${\cal F}_t$ is the fraction of coalescing binaries
that have experienced a triple interaction. This can be estimated simply by knowing
the likelihood of forming triple systems because of two subsequent 
mergers. The galaxy merger rate drops dramatically 
at low redshift, and the typical timescale between two subsequent 
major merger could be as long as $\sim10^{10}$ yr.
This means that massive galaxies may have experienced, on average, 
just one major merger since $z=1$ (see, e.g.~\citealt{bell06}).
If we assume survival time of a binary is $\sim10^9$ yrs, adopting 
the simplifying assumptions of uncorrelated mergers with a Poissonian 
delay distribution with a characteristic time of $10^{10}$ yr, 
the probability of having two subsequent mergers in a $10^9$ yr time 
interval is $\sim 0.1$. We will consider two different 
situations, choosing the fraction of SMBH binaries experiencing a triple 
interaction to be ${\cal F}_t=0.1$ or ${\cal F}_t = 0.5$.

\begin{figure}
\resizebox{\hsize}{!}{\includegraphics[scale=1,clip]{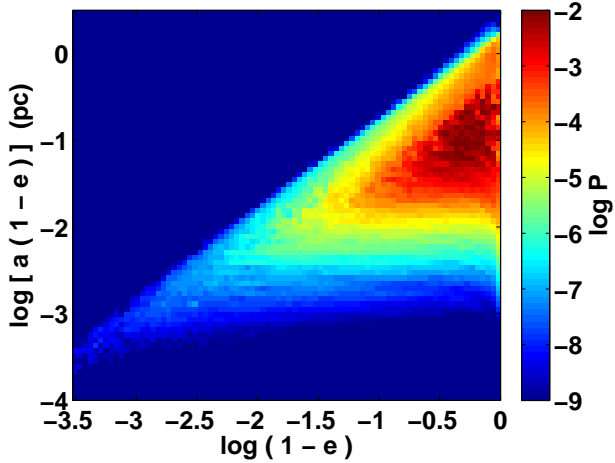}} 
\caption{Two dimensional joint probability distribution ${\cal P}(r_p,e)$ for the 
inner binary in the $[r_p,e]$ plane (where $r_p=a(1-e)$). 
The distribution is obtained averaging over the 1000 3-body experiments 
described in Section~\ref{3body}
\label{fig6} } 
\end{figure}

By knowing ${\cal F}_t$, we can write the coalescence rate of binaries 
that have experienced a triple interactions as ${\cal F}_t\times{d^3N/dzd{\cal M} dt_r}$. 
From the 3-body scattering presented in Section~\ref{3body}, we derive 
the joint probability distribution for the inner binary
of having a certain periastron and a certain eccentricity, ${\cal P}(r_p,e)$.
This quantity is plotted in figure \ref{fig6} for our set of the 1000 3-body 
realizations. This probability distribution refers to systems with mean total
mass $\sim 4\times10^8\msun$. To extend it to a wider range of masses, 
we assume a triplet lifetime ${\cal T}=10^9$ yrs independently of the 
masses (which are in a narrow range peaked around
$10^8\msun$ in our case) and we rescale ${\cal P}(r_p,e)$ so that, in the GW dominated
regime, elements in the $(r_p, e)$  space having the same coalescence
timescale $T_{\rm gw}(r_p,e)$, have the same probability value 
${\cal P}(r_p,e)$. Since $T_{\rm gw}\propto a^4/[M_1M_2(M_1+M_2)]$, assuming
an invariant binary mass ratio distribution in the relevant mass range
(which is a good approximation given the narrow mass range we are dealing with), 
the $y$ axis in figure \ref{fig6} is rescaled for any given total mass 
of the binary $M$ according to $(M/4\times10^8\msun)^{3/4}$. We then compute 
the distribution of eccentric binaries emitting an observable burst as:
\begin{align}
N({\cal M},z,r_p,e) & = &\frac{d^3N}{dzd{\cal M} dt_r}\times{\cal F}_t\times{\cal T}\times{\cal P}(r_p,e)\times \nonumber\\
&& \times{\rm min}[1,(T_{\rm obs}/T_{\rm orb})].
\label{ntrip}
\end{align}
Where the factor ${\rm min}[1,(T_{\rm obs}/T_{\rm orb})]$ takes into
account the fact that if the binary period is longer than the 
observation time, only a fraction $T_{\rm obs}/T_{\rm orb}$ of the
systems is actually bursting during the observation. 

\subsection{Practical computation of the signal}
The relevant frequency band for pulsar timing observations is between 
$1/T_{\rm obs}$ and the Nyquist frequency 
$1/(2\Delta t)$ -- where $\Delta t$ is the time between two adjacent observations--, 
corresponding to $3\times 10^{-9}$ Hz - $10^{-7}$ Hz. The frequency resolution bin 
is $\Delta f=1/T_{\rm obs}$, and we assume $T_{\rm obs}=10$ yr throughout the paper. 
Every realistic frequency--domain computation of the signal has to take into account 
the frequency resolution bin $\Delta f$ of the observation. The signal is therefore 
evaluated for discrete frequency bins $\Delta f_j$ centered at discrete values 
of the frequency $f_j$, where $f_{(j+1)}=f_j+\Delta f$. 
What we actually collect in our code is the numerical distribution 
$\Delta^3N/\Delta z\Delta{\cal M} \Delta f_r$, where $\Delta f_r=(1+z)\Delta f$.
The integral in equation (\ref{hch2}) is then replaced as a sum over redshift and chirp mass,
and the value of the characteristic strain at each discrete frequency $f_j$ is computed as
\begin{equation}
h_c^2(f_j) =\sum_{z}\sum_{{\cal M}}\, \frac {\Delta^3N}{\Delta z\Delta{\cal M} \Delta f_{r,j}}f_r\,
h^2(z,{\cal M},f_r){\Delta z}{\Delta {\cal M}},
\label{hch2num}
\end{equation}
where $\Delta f_{r,j}=(1+z)\Delta f_j$ is the $j$th frequency bin shifted according to the cosmological
redshift of the sources. Equation (\ref{hch2num}) is simply read as the sum of the squares of the 
characteristic strains of all the sources emitting in the {\it observed} frequency bin $\Delta f_j$.
If we produce a family of $\alpha=1,...,K$ sources by performing a Monte Carlo sampling 
of the numerical distribution $\Delta^3N/\Delta z\Delta{\cal M} \Delta f_r$
of the emitting binary population, the characteristic strain is computed as
\begin{equation}
h_c^2(f_j) =\sum_{\alpha=1}^K h_{c,\alpha}(z,{\cal M}, f_{\alpha,r})^2\Theta[f_{\alpha,r},\Delta f_j(1+z)]
\label{hch2num2}
\end{equation}
where $\Theta[f_{\alpha,r},\Delta f_j(1+z)]=1$ if $f_{\alpha,r} \in \Delta f_j(1+z)$  and is null elsewhere.
To recover equation (\ref{hch2num}), the {\it characteristic amplitude} of the individual source is given by: 
$h_{c,\alpha}^2=h_{\alpha}^2f_{\alpha,r}/\Delta f_{r,j}\approx h_{\alpha}^2f_j/\Delta f_{j}=h_{\alpha}^2f_jT_{\rm obs}$; 
\i.e., the sky and polarization averaged amplitude square, multiplied by the number of 
cycles completed in the observation time. The induced rms residual of each individual source is then given
by equation (\ref{e:deltatgw}). Note that in the limit of large $K$
(formally, $K\rightarrow\infty$), $h_c(f_j)$ computed according to equation (\ref{hch2num2}) 
is independent of $T_{\rm obs}$ (because the increment of the contribution of each single source according 
to the number of cycles completed during $T_{\rm obs}$ is balanced by the fact that we sum over a 
frequency bin that is proportional to $1/T_{\rm obs}$), and its value coincides with the one obtained
from the standard energy based definition of $h_c(f)$ \citep{sesana08}. On the other hand,
when $K$ is small (i.e. we sum over a small number of sources), 
fluctuations become important in the computation of the signal
in each frequency bin. Numerical computation according to 
equation (\ref{hch2num2}) allows us to account for signal fluctuations, 
which are missing in the analytical definition of the characteristic amplitude 
of the GW spectrum \citep[e.g.][]{phinney01}, but are important in the actual computation of the observed signal.
Given $h_c$, the induced rms timing residual produced by the whole emitting population 
is simply given by $h_c(f_i)/(2\pi f_i)$.

We generate a population of emitting binaries according to the numerical distribution 
$\Delta^3N/\Delta z\Delta{\cal M} \Delta f_r$,
and we sum all the $h_{c,\alpha}$ contributions in every frequency bin to obtain 
the characteristic strain of the signal. 
We then generate, again using a Monte--Carlo sampling, a population of emitting eccentric binaries in 
triple systems from the distribution given in equation (\ref{ntrip}) and we compute their GW bursts and the induced rms 
residuals according to equations (\ref{hrms}, \ref{hobs}, \ref{e:deltatgws}, \ref{e:deltatgw2}). 
For the few systems reaching $10^{-5}$ Hz with their higher harmonics, we also compute the SNR produced in the 
{{\it LISA}} detector using equation (\ref{snr}), adopting the $S_{f}$ given in equation (48) 
of~\citet{barack04}, extended downward to $10^{-5}$ Hz as described in Section~\ref{obsquants}.
We consider five different SMBH binary populations presented in SSV09 (Tu-SA, Tu-DA, Tu-NA, La-SA, Tr-SA)
with two different fractions of triplets ${\cal F}_t=0.1, 0.5$, for a grand total of 10 
different models. We run 50 (when ${\cal F}_t=0.5$; 100 if ${\cal F}_t=0.1$) 
independent Monte--Carlo realizations of each single model, which allows 
us to perform a statistical study of the properties of the bursting sources. 
We consider only systems with $e>0.66$, because highly eccentric systems are those expected to burst 
at high frequencies, where the contribution of the overall circular binary population declines. 
And also because high eccentricities result in a well defined burst shape 
which may be essential to distinguish it from periodic sources. 

\section{Results}

\subsection{description of the signal}
\begin{figure}
\resizebox{\hsize}{!}{\includegraphics[scale=1,clip]{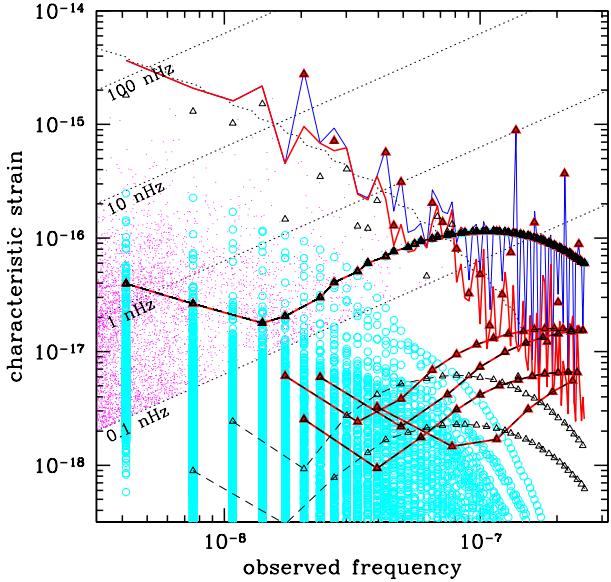}} 
\caption{Representation of all the relevant features of a Monte--Carlo generated signal. The 
Tu-DA model with ${\cal F}=0.5$ is assumed. The jagged blue line is an 
individual Monte--Carlo realization of the signal. The small black triangles label the characteristic
strain of the brightest source in each frequency bin. If the source is resolvable, it is also labeled with
a big red triangle. The jagged red line is the stochastic level of the signal, i.e., 
once the resolvable sources in each frequency bin are subtracted. The magenta points label all the 
systems producing an rms residual (computed through equation (\ref{e:deltatgw})) 
larger than 0.1ns over 10 years. The cyan 'arcs' of dots, 
represent the contribution to the signal coming from eccentric binaries in triple systems
(bursting sources), again assuming that their total rms residual is larger than 0.1ns
(equation (\ref{e:deltatgw2})). The arc-like black (red) tracks represent the spectrum of the 
more luminous (resolvable) bursting systems in the realization, and have the only purpose of 
guiding the reader eye. The dotted oblique lines mark different rms residual levels as a function of the frequency.  
\label{fig7} } 
\end{figure}

\begin{figure*}
\resizebox{\hsize}{!}{\includegraphics[scale=1,clip]{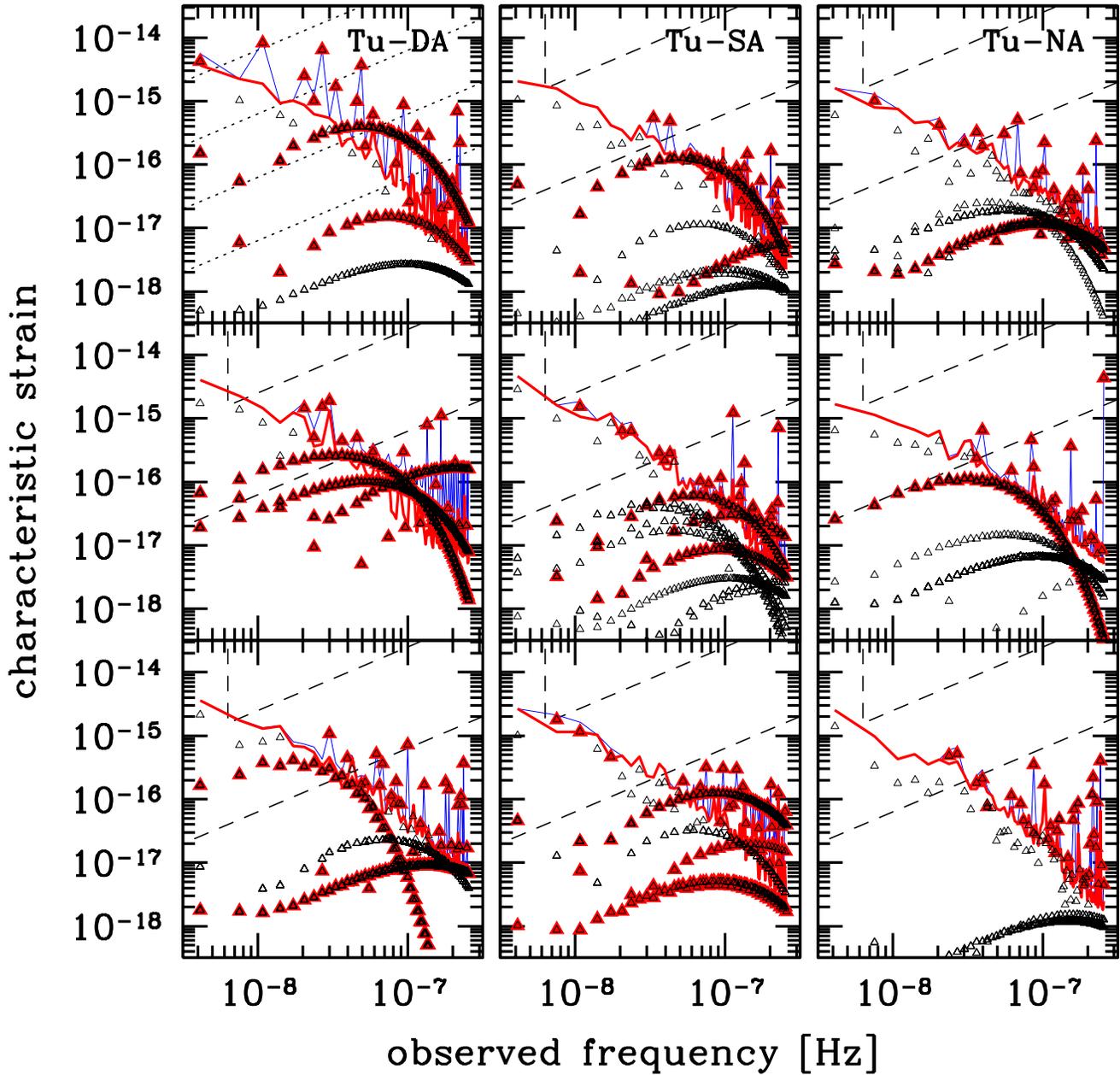}} 
\caption{Sample of individual Monte--Carlo realizations of the signal generated using Tu-DA (left panels),
Tu-SA (central panels) and Tu-NA (right panels) models (${\cal F}=0.5$). Line and point style as in figure \ref{fig7}.
The two dashed lines in each panel represent the sensitivity of the PPTA (upper) survey and an 
indicative sensitivity of 1ns for SKA (lower).  
\label{fig8} } 
\end{figure*}

All the relevant features of the signal are plotted in figure \ref{fig7} for 
a realization of the Tu-DA model with  ${\cal F}_t=0.5$. A Monte--Carlo generated signal  
is depicted as a blue jagged line. The magenta points represent all the 
binary systems producing a $\delta t_\mathrm{gw}>0.1$ ns; 
there are $\sim 4000$ sources in this particular realization. The cyan 'arcs' of dots, 
represent the contribution to the signal coming from eccentric binaries in triple systems
(bursting sources), where contributions from all harmonics falling in the same frequency bin were added in quadrature.
The black triangles correspond to the brightest source in each frequency bin. And if a 
source is brighter than the sum of all the contributions coming from the other sources emitting
in the same bin, we consider that source {\it resolvable} and we mark it with a superposed red triangle.
The red jagged line is the resulting {\it stochastic level} of the signal,
after the contribution from the resolvable sources has been subtracted. 
The arc-like black (red) tracks represent the more luminous (resolvable) bursting
systems in the realization. In this particular case there were five resolvable bursts with
rms residual $\delta t_{\rm gw}=3.5, 0.07, 0.04, 0.01, 0.002$ ns. However, considering  
realistic PTA sensitivities achievable in the near future ($\sim1$ns, with the SKA), 
only the brightest one would have a good chance of being detected.

We note that we introduced
the concept of {\it resolvable source} in the frequency domain, assuming that a source 
is resolvable if its strain is larger than the 
sum of the strains of all the other sources in that frequency bin. This definition is, however, 
only appropriate for monochromatic sources. A very eccentric burst, emitting a whole spectrum of harmonics, may
not be the brightest source in any of the frequency bins, however, it may produce a significantly larger
rms residuals with respect to other individual circular binaries. Moreover, in the time domain, 
the signature of these bursts is quite different with respect to periodic circular binaries, 
resulting in long bumps or narrow well localized bursts (see figure \ref{fig11}). 
Given these caveats, we will also present results 
in terms of total number of sources, independent of their resolvability according to our definition.

A sample of different realizations of the signal is collected in figure \ref{fig8},
for individual realizations of the three different Tu models. 
Only the brightest sources are plotted in this case. 
Given the small number of systems involved, their phenomenology is quite 
variable. For example, the realization illustrated in the left-middle panel,
shows three resolvable bursts with $\delta t_{\rm gw}\gtrsim3$ns; the one in the lower-
left panel, does not show any individually resolvable bursts.

\subsection{Statistic of bursting sources}

\begin{figure}
\resizebox{\hsize}{!}{\includegraphics[scale=1,clip]{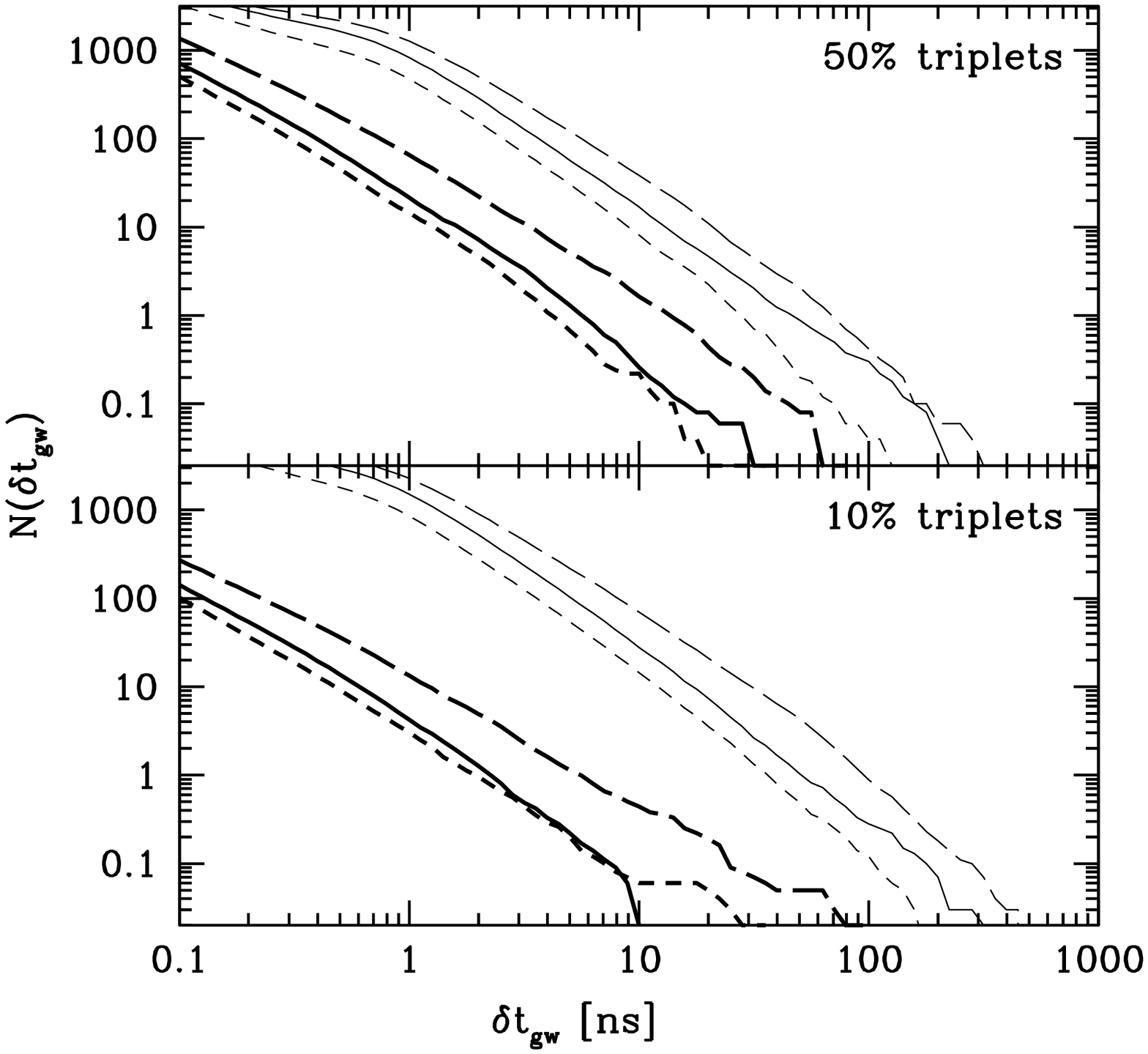}} 
\caption{Cumulative number $N(\delta t_\mathrm{gw})$ of circular binaries (thin lines) and bursting
triplets (thick lines) emitting over a given $\delta {t_{\rm GW}}$ threshold as a function 
of $\delta {t_{\rm GW}}$. In each panels the different linestyles refer to the
Tu-SA (solid), Tu-DA (long--dashed) and Tu-NA (short--dashed) models. The fraction of
triplets assumed is labeled in each panel.  
\label{fig9} } 
\end{figure}

\begin{figure}
\resizebox{\hsize}{!}{\includegraphics[scale=1,clip]{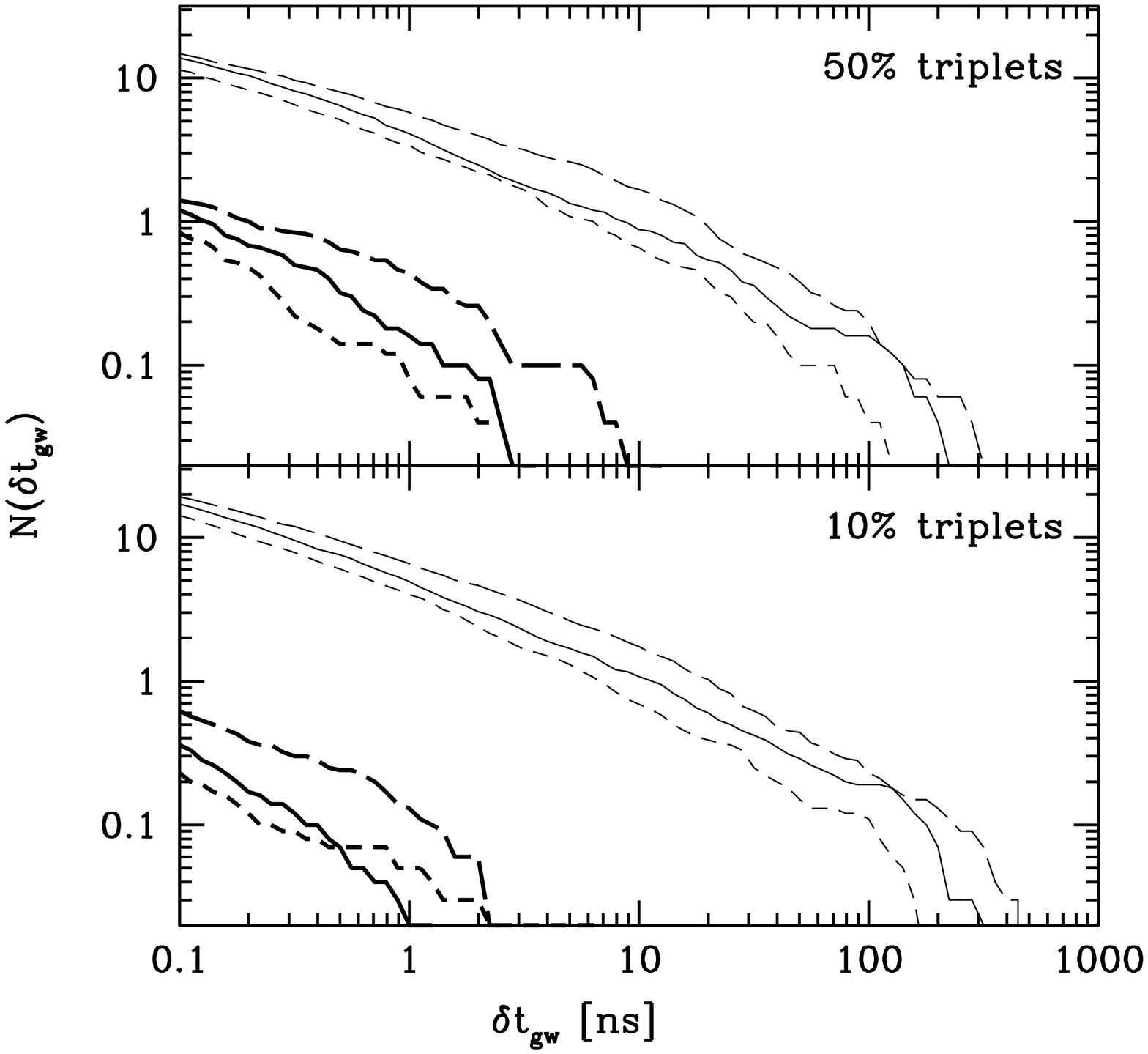}} 
\caption{Same as figure \ref{fig9}, but considering only the resolvable sources in the computation
of $N(\delta t_\mathrm{gw})$ according to equation (\ref{e:N}). Linestyle as in figure \ref{fig9}. 
\label{fig10} } 
\end{figure}

To quantify the statistics of the bursting sources, we cast the results in terms of the 
cumulative number of sources as a function of the timing residuals:
\begin{equation}
N(\delta t_\mathrm{gw}) = \int_{\delta t_\mathrm{gw}}^\infty 
\frac{d N}{d(\delta t_\mathrm{gw}')}d(\delta t_\mathrm{gw}')\,,
\label{e:N}
\end{equation}
where the distribution $d N/d(\delta t_\mathrm{gw}')$ is the average over the 50 (100) Monte--Carlo
realizations of each model. We compute this average both considering all the sources emitting over a given
$\delta t_\mathrm{gw}$ threshold (obtaining the total distribution of bursting sources), 
as well as considering only resolvable sources as defined in the previous section
(obtaining the distribution of bursting resolvable sources).  In figure \ref{fig9}, $N(\delta t_\mathrm{gw})$ for all the sources is shown.
Depending on the adopted model, and on the fraction of triplets assumed, there are few hundred to 
few thousand binaries contributing to the signal at a level $\gtrsim 1$ns. The number of 
triplets over this threshold is between 20 and 60 assuming ${\cal F}=0.5$, and, not surprisingly,
a factor of 5 lower if we assume ${\cal F}=0.1$. If triple interactions of SMBHs are common
(say, ${\cal F}>0.1$), we may therefore expect 1-to-100 bursts from eccentric sources contributing 
to the GW signal at a residual level of $>1$ns. The eccentricity distribution of these bursts
is basically flat in the considered eccentricity range $(0.66,1)$. If we consider {\it resolvable sources}
only, the figures are not as promising. As shown in figure \ref{fig10}, a timing precision
of 0.1 ns is needed to guarantee the detectability of a resolvable burst if ${\cal F}=0.5$. 
At a 1 ns level, we have less than one resolvable burst, we can then interpret the results in 
terms of the probability of having such bursts in our observable Universe. This probability
ranges from 2\% to 50\% depending on the adopted model, and the eccentricity distribution
of these resolvable events is biased towards high values, peaking around $e=0.9$.
La-SA and Tr-SA give similar results both qualitatively and quantitatively, therefore
we don't plot them in the figures in order to keep them clear.
Again, we stress the fact that our definition of {\it resolvable source} is rather arbitrary, 
and does not take into account for the peculiar shape of the burst, we then consider these 
figures as lower limits to the actual detectability of these bursts.
  
\subsection{Signal samples in the time domain}
To give a feeling of how the actual signals would appear, we also computed residuals in the
time domain for selected sources. To this purpose, we evolved the system using 
equations (27)-(31) of~\citet{barack04} assuming non spinning SMBHs. We then computed 
all the components $ h_{+,n}$ and $ h_{\times,n}$ (following~\citealt{pierro01}) and we
finally evaluated the residuals $R(T)$ integrating equation (\ref{restd}). 
The actual shape of the residuals is rather complex and depends on the geometry 
of the system: the relative orientation of the source to the pulsars (encoded in the 
direction cosines $\alpha$, $\beta$, and $\gamma$ in equation (\ref{restd})); the 
polarization angle of the source $\Psi$; the inclination $i$; the initial phase of 
the orbit $\Phi_0$; and an angle $\phi_p$ describing the orientation of the periastron 
in the orbital plane (see, e.g.,~\citealt{barack04} for a definition of all these
quantities).

\begin{figure}
\resizebox{\hsize}{!}{\includegraphics[scale=1,clip]{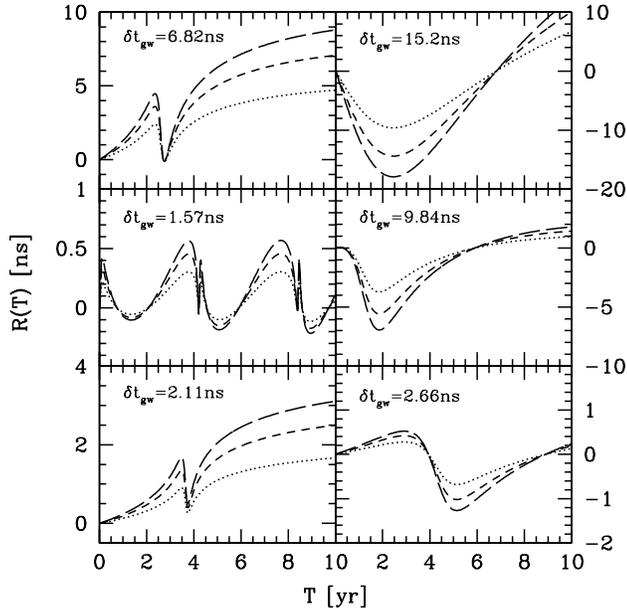}} 
\caption{Examples of timing residual, found in a particular realization of the
model Tu-DA, computed according to equation (\ref{restd}). Different linestyles 
correspond to different aperture angle $\theta$ between the pulsar and the source; 
$\theta=\pi/2$ (dotted), $\pi/3$ (short--dashed), $\pi/6$ (long--dashed). In all
the cases $\Psi=\pi/4$ and $i=\pi/3$. $\phi_p$ is random and $\Phi_0$ is 
chosen so that the burst occurs during the observation. 
The rms residual computed according to equation (\ref{e:deltatgw2}) are also shown. 
Parameters of the sources are listed in table \ref{tab.2}.
\label{fig11} } 
\end{figure}

Examples of the phenomenology of bursting sources are given in figure \ref{fig11} for
a sample of eccentric systems found in one selected realisation of the model Tu-DA. 
In the left panels we show the three brightest {\it resolvable} sources, while in the
right panels we show three of the brightest bursts which would be {\it unresolvable} according
to our definition, because their power spectra would be overwhelmed by the signal
produced by the standard circular binaries found in the realization. Parameters of the binaries
are given in table \ref{tab.2}. Bursts can be generated by very eccentric-long period 
binaries (as in the two lower panels), or by relatively short-period systems (e.g., central
left panel), in which case multiple bursts are visible in the observation times. 
The width of the burst depends on the periastron passage timescale: systems with $T_p\ll T_{\rm obs}$
produce narrow features in the data stream (e.g., lower left panel), while systems 
with $T_p\approx T_{\rm obs}$ give a characteristic bump shaping all over the data span (e.g. upper 
right panel). Given the integral nature of the signal (equation (\ref{restd})), its shape 
is also heavily dependent on $\Phi_0$ (e.g. the cumulative residual can be positive or negative
depending on the binary orbital phase at the beginning of the detection), on $\phi_p$ and on $\Psi$.
The inclination of the source $i$ and its aperture angle to the pulsar, determine
the amplitude of the signal.

\begin{table}  
\begin{center}
\begin{tabular}{cccccc} \hline 
$M_1$ [M$_\odot$] & $M_2$ [M$_\odot$] & $f_r$ [Hz] & $z$ & $e$ & $\delta t_{\rm gw}$[ns]\\ 
\hline 
$1.4\times10^9$ & $4.4\times 10^8$ & $1.92\times10^{-10}$ & 0.965 & 0.232 & 6.82\\ 
$9.9\times10^8$ & $4.2\times 10^7$ & $7.5\times10^{-9}$ & 0.88 & 0.082 & 1.57\\ 
$9.8\times10^8$ & $3.9\times 10^8$ & $1.38\times10^{-10}$ & 0.979 & 0.775 & 2.11\\ 
$6.8\times10^8$ & $4.1\times 10^8$ & $2.31\times10^{-11}$ & 0.973 & 0.922 & 2.66\\ 
$1.2\times10^9$ & $4.1\times 10^8$ & $4.03\times10^{-10}$ & 0.84 & 0.239 & 9.84\\ 
$5.6\times10^9$ & $2.7\times 10^8$ & $2.11\times10^{-10}$ & 0.75 & 0.086 & 15.2\\ 
\hline 
\end{tabular} 
\end{center} 
\caption{Parameters of the sources plotted in figure \ref{fig11}. Rows
in the table (from the top to the bottom), correspond to panels of figure \ref{fig11}
considered counterclockwise, starting from the upper left panel.\label{tab.2}}
\end{table}

\subsection{A note for {{\it LISA}}}
We also collected catalogs of systems bursting in the {{\it LISA}} window, to check
for detectability. Unfortunately, prospects for detection with {{\it LISA}} are 
not as promising as for PTAs. In a total of 750 realization of the ten different
models, we found $\sim 50$ sources bursting in the {{\it LISA}} window producing 
an SNR$>0.1$. Unfortunately, none of them  produced a SNR$>8$, necessary for a 
confident detection. We then conclude, that even with a consistent population of SMBH triplets
forming during the cosmic history, burst from massive (say, ${\cal M}\sim 10^8\msun$) eccentric
binaries are unlikely to be produced at a significant rate for {{\it LISA}}. On the other hand,
if formation of triple systems was common in the past, for system in the {{\it LISA}} mass range
($\sim 10^5-10^7\msun$), very peculiar signals from coalescing eccentric binaries may be 
common in the data stream. However, this is beyond the scope of the present
paper, where we focused on massive binaries (${\cal M}> 10^7\msun$) only.

\section{Conclusions}
\label{conclusions}

We have addressed in this work three different points in the evolution of
triplets of SMBHs in the Universe: The Astrodynamics of the system,  the
potential GW signature and the detectability.

We have performed eight different direct-summation $N-$body  simulations, one
including more than half a million of particles, to calibrate
1,000 3-body scattering experiments, which include post-Newtonian
corrections, in order to have a statistical description of the system. Both
numerical tools agree that the inner binary of SMBHs will go through a phase of
extremely high eccentricity, which is the motivation for the rest of the work.

These three-body excitations of episodic high eccentricity  configurations of
the close SMBH binary produce interesting GW bursts that may be detectable with
forthcoming experiments such as PTAs and {{\it LISA}}.  The extreme
eccentricities of such bursts on one hand would leave a very distinctive
signature, but on the other require the development of appropriate analysis
techniques.

To compute likely event rates, we extracted catalogues of merging galaxies
from the Millennium Run, and we populated them with SMBHs following the known
MBH-bulges relations. We then estimated the fractions of triplets and their
eccentricity distribution and we computed the induced signals in both PTAs and
the {{\it LISA}} detector.

We found that, depending on the details of the SMBH population model, if 
the fraction of triplets is $\ge 0.1$, few to a hundred of GW bursts would 
be produced at a $>1$ ns level in the PTA frequency domain. Most
of the signals will be washed out in the confusion noise due to the emission
of `ordinary' low eccentric binaries. However, their peculiar features may
guide the development of targeted data analysis techniques, that may help
to recognize them even if overwhelmed by the confusion noise.
Employing a minimal criterion for source resolvability (which provides
a strict lower limit), we found that less than one system may be actually
pinned down at ns precisions. By running several dozens of Monte Carlo
realization of the signal from the cosmological population of SMBH binaries 
and triplets we quantified a statistical $2-50$\% chance of having a 
resolvable burst in the Universe (assuming 10 yrs of observation).
The probability for detection with {\rm {\it LISA}} is essentially nil.
However, we stress the fact that we focused on systems with ${\cal M}>10^7\msun$; 
our results then simply imply that it is extremely unlikely that a 
system which would normally emit outside the {\it LISA} range will 
produce a burst in the {\it LISA} window because of resonant three body 
interactions. On the other hand, if a consistent fraction of {light}
binaries (${\cal M}<10^7\msun$) is involved in triple systems, we may expect 
several eccentricity-driven coalescences to be observed by {\it LISA}. This
eventuality would call for the development of extremely eccentric templates
($e>0.9$) for merging SMBHs, and of adequate analysis techniques to 
extract the signal.

\section*{Acknowledgments}
The work of PAS has been supported by the Deutsches Zentrum f{\"u}r Luft- und
Raumfahrt. PAS is indebted with Sterl Phinney for discussions which motivated
the work and with Marc Freitag for comments on the article and help with some
of the diagrams.  PAS, AS and MB acknowledge the support of the Aspen Center
for Physics. Some of the numerical simulations were done with the {\sc
Tuffstein} cluster located at the Max-Planck Institut f{\"u}r
Gravitationsphysik (Albert Einstein-Institut). RS and PAS acknowledge computing
time on the GRACE cluster in Heidelberg (Grants I/80 041-043 of the Volkswagen
Foundation and 823.219-439/30 and /36 of the Ministry of Science, Research and
the Arts of Baden-W{\"u}urttemberg). MB acknowledges the support of NASA grant
NNX08AB74G and the Center for Gravitational Wave Astronomy, supported by NSF
award 0734800.

\label{lastpage}

\end{document}